\documentclass[reprint,amsmath,amssymb,
aps,prl,lengthcheck,nobibnotes,superscriptaddress]{
revtex4-2}

\usepackage{graphicx}
\usepackage{dcolumn}
\usepackage{bm}
\usepackage{hyperref}
\usepackage[geometry]{ifsym}
\usepackage{color}
\begin{document}


\title{Beta delayed neutron emission of $N=84$ $^{132}$Cd  }


%

\author{M. Madurga}
\affiliation{Dept. of Physics and Astronomy, University of Tennessee, Knoxville, Tennessee 37996, USA}
\author{ Z.Y. Xu}
\affiliation{Dept. of Physics and Astronomy, University of Tennessee, Knoxville, Tennessee 37996, USA}
\author{ R. Grzywacz} 
\affiliation{Dept. of Physics and Astronomy, University of Tennessee, Knoxville, Tennessee 37996, USA}
\author{ M.R. Mumpower }
\affiliation{Obsidian Research, Fort Wayne, IN 46835, USA}
\affiliation{Department of Physics and Astronomy, University of Notre Dame, Notre Dame, IN 46556, USA}
\affiliation{Computational and Artificial Intelligence Division, Los Alamos National Laboratory, Los Alamos, NM 87545, USA}
\author{A.N. Andreyev}
\affiliation{School of Physics, Engineering and Technology, University of York, North Yorkshire YO10 5DD, United Kingdom}
\author{G.	Benzoni}
\affiliation{Dipartimento di Fisica and INFN, Sezione di Milano, Milano, Italy}
\author{M.J.G.	Borge}
\affiliation{CERN, CH-1211 Geneva 23, Switzerland}
\author{C.	Costache}
\affiliation{“Horia Hulubei” National Institute of Physics and Nuclear Engineering, RO-077125 Bucharest, Romania}
\author{I. Cox}
\affiliation{Dept. of Physics and Astronomy, University of Tennessee, Knoxville, Tennessee 37996, USA}
\author{S. Cupp}
\affiliation{Theoretical Division, Los Alamos National Laboratory, Los Alamos, NM 87545, USA}
\author{B.	Dimitrov}
\affiliation{Institute for Nuclear Research and Nuclear Energy, BAS, Sofia, Bulgaria}
\author{P. Van Duppen}
\affiliation{Instituut voor Kern- en Stralingsfysica, K.U. Leuven, BE-3001 Leuven, Belgium}
\author{L.M. Fraile}
\affiliation{Grupo de Física Nuclear, EMFTEL \& IPARCOS, Universidad Complutense de Madrid, E-28040 Madrid, Spain}
\affiliation{CERN, CH-1211 Geneva 23, Switzerland}
\author{S.	Franchoo}
\affiliation{Université Paris-Saclay, CNRS/IN2P3, IJCLab, 91405 Orsay, France}
\author{H. Fynbo}
\affiliation{Department of Physics and Astronomy, Aarhus University, DK-8000 Aarhus C, Denmark}
\author{B.	Gonsalves}
\affiliation{Department of Physics, P.O. Box 64, FIN-00014 University of Helsinki, Finland}
\author{A.	Gottardo}
\affiliation{Université Paris-Saclay, CNRS/IN2P3, IJCLab, 91405 Orsay, France}
\author{P.T. Greenless}
\affiliation{Department of Physics, P.O. Box 64, FIN-00014 University of Helsinki, Finland}
\author{A. Gross}
\affiliation{Theoretical Division, Los Alamos National Laboratory, Los Alamos, NM 87545, USA}
\author{C.J.	Gross}
\affiliation{Physics Division, Oak Ridge National Laboratory, Oak Ridge, Tennessee 37830, USA}
\author{L.J. Harkness-Brennan}
\affiliation{Department of Physics, Oliver Lodge Laboratory, University of Liverpool, Liverpool L69 7ZE, United Kingdom}
\author{M. Huyse}
\affiliation{Instituut voor Kern- en Stralingsfysica, K.U. Leuven, BE-3001 Leuven, Belgium}
\author{D.S. Judson}
\affiliation{Department of Physics, Oliver Lodge Laboratory, University of Liverpool, Liverpool L69 7ZE, United Kingdom}
\author{S.	Kisyov}
\affiliation{“Horia Hulubei” National Institute of Physics and Nuclear Engineering, RO-077125 Bucharest, Romania}
\author{K.	Kolos}
\affiliation{Lawrence Livermore National Laboratory, Livermore, California 94550, USA}
\author{J. Konki}
\affiliation{Department of Physics, P.O. Box 64, FIN-00014 University of Helsinki, Finland}
\author{J. Kurcewicz}
\affiliation{CERN, CH-1211 Geneva 23, Switzerland}
\author{I. Lazarus}
\affiliation{STFC Daresbury, Daresbury, Warrington WA4 4AD, United Kingdom}
\author{R.	Lic\u{a}}
\affiliation{CERN, CH-1211 Geneva 23, Switzerland}
\author{L.	Lynch}
\affiliation{CERN, CH-1211 Geneva 23, Switzerland}
\author{M. Lund}
\affiliation{Department of Physics and Astronomy, Aarhus University, DK-8000 Aarhus C, Denmark}
\author{N. Marginean}
\affiliation{“Horia Hulubei” National Institute of Physics and Nuclear Engineering, RO-077125 Bucharest, Romania}
\author{R. Marginean}
\affiliation{“Horia Hulubei” National Institute of Physics and Nuclear Engineering, RO-077125 Bucharest, Romania}
\author{C. Mihai}
\affiliation{“Horia Hulubei” National Institute of Physics and Nuclear Engineering, RO-077125 Bucharest, Romania}
\author{I.	Marroquin}
\affiliation{Instituto de Estructura de la Materia, CSIC, E-28040 Madrid, Spain}
\author{C.	Mazzocchi}
\affiliation{Faculty of Physics, University of Warsaw, Warszawa PL 00-681, Poland}
\author{D.  Mengoni}
\affiliation{INFN Laboratori Nazionali di Legnaro, Padova, Italy}
\author{A.I Morales}
\affiliation{Instituto de Física Corpuscular, CSIC Universidad de Valencia, E-46071 Valencia, Spain}
\author{E. Nacher}
\affiliation{Instituto de Física Corpuscular, CSIC Universidad de Valencia, E-46071 Valencia, Spain}
\author{A. Negret}
\affiliation{“Horia Hulubei” National Institute of Physics and Nuclear Engineering, RO-077125 Bucharest, Romania}
\author{R.D. Page}
\affiliation{Department of Physics, Oliver Lodge Laboratory, University of Liverpool, Liverpool L69 7ZE, United Kingdom}
\author{S. Pascu}
\affiliation{“Horia Hulubei” National Institute of Physics and Nuclear Engineering, RO-077125 Bucharest, Romania}
\author{S.V. Paulauskas}
\affiliation{Dept. of Physics and Astronomy, University of Tennessee, Knoxville, Tennessee 37996, USA}
\author{A.	Perea}
\affiliation{Instituto de Estructura de la Materia, CSIC, E-28040 Madrid, Spain}
\author{M.	Piersa-Si\l{}kowska}
\affiliation{Dept. of Physics and Astronomy, University of Tennessee, Knoxville, Tennessee 37996, USA}
\affiliation{Grupo de Física Nuclear, EMFTEL \& IPARCOS, Universidad Complutense de Madrid, E-28040 Madrid, Spain}
\affiliation{Faculty of Physics, University of Warsaw, Warszawa PL 00-681, Poland}
\author{V. Pucknell}
\affiliation{STFC Daresbury, Daresbury, Warrington WA4 4AD, United Kingdom}
\author{P. Rahkila}
\affiliation{Department of Physics, P.O. Box 64, FIN-00014 University of Helsinki, Finland}
\author{E. Rapisarda}
\affiliation{CERN, CH-1211 Geneva 23, Switzerland}
\author{F. Rotaru}
\affiliation{“Horia Hulubei” National Institute of Physics and Nuclear Engineering, RO-077125 Bucharest, Romania}
\author{C.	Sotty}
\affiliation{Instituut voor Kern- en Stralingsfysica, K.U. Leuven, BE-3001 Leuven, Belgium}
\author{S. Taylor}
\affiliation{Dept. of Physics and Astronomy, University of Tennessee, Knoxville, Tennessee 37996, USA}
\author{O.	Tengblad}
\affiliation{Instituto de Estructura de la Materia, CSIC, E-28040 Madrid, Spain}
\author{V.	Vedia}
\affiliation{Grupo de Física Nuclear, EMFTEL \& IPARCOS, Universidad Complutense de Madrid, E-28040 Madrid, Spain}
\author{D.	Verney}
\affiliation{Université Paris-Saclay, CNRS/IN2P3, IJCLab, 91405 Orsay, France}
\author{R. Wadsworth}
\affiliation{School of Physics, Engineering and Technology, University of York, North Yorkshire YO10 5DD, United Kingdom}
\author{N. Warr}
\affiliation{Department of Physics, Oliver Lodge Laboratory, University of Liverpool, Liverpool L69 7ZE, United Kingdom}
\affiliation{Institut f\"ur Kernphysik, Universit\"at zu K\"oln, 50937 K\"oln, Germany}
\author{H. de Witte}
\affiliation{Instituut voor Kern- en Stralingsfysica, K.U. Leuven, BE-3001 Leuven, Belgium}

\date{\today}

\begin{abstract}
Using the time-of-flight technique, we measured the beta-delayed neutron emission of $^{132}$Cd. From our large-scale shell model (LSSM) calculation using the N$^3$LO interaction [Z.Y. Xu et al., Phys. Rev. Lett. 131, 022501 (2023)], we suggest the decay is dominated by the transformation of a neutron in the $g_{7/2}$ orbital, deep below the Fermi surface, into a proton in the $g_{9/2}$ orbital. We compare the beta-decay half-lives and neutron branching ratios of nuclei with $Z<50$ and $N\geq82$ obtained with our LSSM with those of leading ``global" models such as Finite-Range Droplet Model (FRDM). Our calculations match known half-lives and neutron branching ratios well and suggest that current leading models overestimate the yet-to-be-measured half-lives. Our model, backed by the $^{132}$Cd decay data presented here, offers robust predictive power for nuclei of astrophysical interest such as {\it r}-process waiting points.


\end{abstract}

\maketitle

{\it Introduction.} The {\it r}-process sensitivity studies \cite{Mumpower2016,Mumpower2024} have identified nuclei southeast of the doubly magic $Z=50$ $N=82$ $^{132}$Sn as
crucial in calculating the {\it r}-process nucleosynthesis for almost all
astrophysical environments. This is due to the robustness of the
$N=82$ and $Z=50$ shell closures \cite{Jones2010}, which causes a dramatic
discontinuity in proton and neutron separation energy. This feature
forces the {\it r}-process sequence of neutron capture to follow a path
aligned with the $N=82$ shell closure and a sharp turn-off towards more
neutron-rich nuclei at $Z=50$. This also results in the unique sensitivity
of the {\it r}-process to nuclear properties of the nuclei below $^{132}$Sn, even
though the $^{132}$Sn itself is not directly involved \cite{Mumpower2016}.
Despite the apparent simplicity of the nuclei in the vicinity of $^{132}$Sn, the presence of high- and low-{\it j} orbitals and large
proton number to neutron number asymmetry \cite{Nowacki2021} results in the emergence of ``exotic" nuclear phenomena. Some examples are the presence of
isomers \cite{Watanabe2013,Phong2019}, the early onset of collectivity in $Z>50$ nuclei \cite{Allmond2017,Gray2020},
or the unexpectedly short half-lives (compared to leading
calculations using ``global” models such as FRDM \cite{Moller2019} or DF3-CQRPA \cite{Borzov2016}) for $Z<50$, $N\geq82$ nuclei. Lorusso and collaborators \cite{Lorusso2015} show that shorter half-lives result in a robust $A\sim$160 peak in the calculated solar {\it r}-process abundance pattern, compared to a much smaller one using calculated half-lives. Their result highlights the importance of accurate predictions for nuclei where no experimental information is yet available. M\"oller, Pfeiffer and Kratz identified the half-life problem in their seminal work on ``speeding up the {\it r}-process" \cite{Moller2003}. They observed that including first-forbidden (FF) beta decay together with allowed Gamow-Teller (GT) transitions decreased their calculated half-lives. However, their latest calculations still do not align with experimental values \cite{Lorusso2015,Moller2019}, leaving the problem unresolved.

Both half-lives \cite{Lorusso2015} and neutron branching ratios \cite{Hall2021,Phong2022} in the $Z<50$ $N\geq82$ region are of crucial importance to model the astrophysical {\it r}-process correctly \cite{Mumpower2016,Mumpower2024}. The difficulty, mentioned above, of ``global" models in reproducing known half-lives and branching ratios indicates that nuclear structure effects play a much more important role than initially assumed. In our recent investigation of the beta decay of $N=84$ $^{133}$In, we provided the first piece of evidence to tackle this problem \cite{Xu2023}. We showed that the decay strength, and thus the half-life, is dominated by the transformation of a neutron deep below the Fermi surface ($\nu$-$1g_{7/2}\rightarrow\pi$-$1g_{9/2}$). We postulated such transformation should be present and thus dominate the decay of all other $N=84$ isotones. However, testing this hypothesis is hampered by the lack of complete spectroscopic studies. While the beta-delayed gamma spectrum has been studied in $^{133,134,135}$In \cite{Hoff1996,Piersa2019,Piersa2021}, $^{132,133}$Cd \cite{Taprogge2014,Jungclaus2016}, and $^{130}$Ag \cite{Kautzsch2000}, no beta-delayed neutron spectroscopy has been done except for $^{133,134}$In \cite{Hoff1996,Hoff2000,Xu2023,Heideman2023}. Beta-delayed neutron spectroscopy has proved to be crucial, as the energy required to transform a neutron deep below the Fermi surface places the dominating transition well above the neutron separation energy \cite{Xu2023}. Only by studying the beta-delayed neutron emission of $Z<50$ $N=82,84$ isotones can we detect the crucial fingerprint, the decay of a neutron deep below the Fermi surface, needed to validate nuclear models in the region.    

In this Letter we report the measurement of the beta-delayed gamma, and for the first time, neutron emission of $^{132}$Cd. From the lack of observation of gamma transitions between states in $^{132}$In, we deduce a neutron branching ratio of 100\%, consistent with Taprogge et al. \cite{Taprogge2014} and Phong et al. \cite{Phong2022}. As the beta-decay populates only neutron-unbound states, we obtained the decay strength distribution from the neutron branching ratios. We calculated the $^{132}$Cd decay strength using a large scale shell model (LSSM) and the N$^3$LO interaction, resulting in a good match to the experimental distribution. We tested our LSSM by calculating the half-lives of $N=82,84$ isotones with atomic number between 43 and 49 (waiting points of the astrophysical {\it r}-process), obtaining values as much as twice as short as the generally used predictions by FRDM-QRPA. We also calculated the neutron branching ratios of $N=82$ to $N=85$ cadmium isotopes, obtaining values systematically larger than ``global" models.  This highlights the importance of validating theoretical calculations with a variety of observables, and shows much work still needs to be done in reliably predicting the properties of nuclei relevant to the {\it r}-process \cite{Mumpower2016,Mumpower2024}.

\begin{figure}
  \begin{center}
  \includegraphics[width=3.5in]{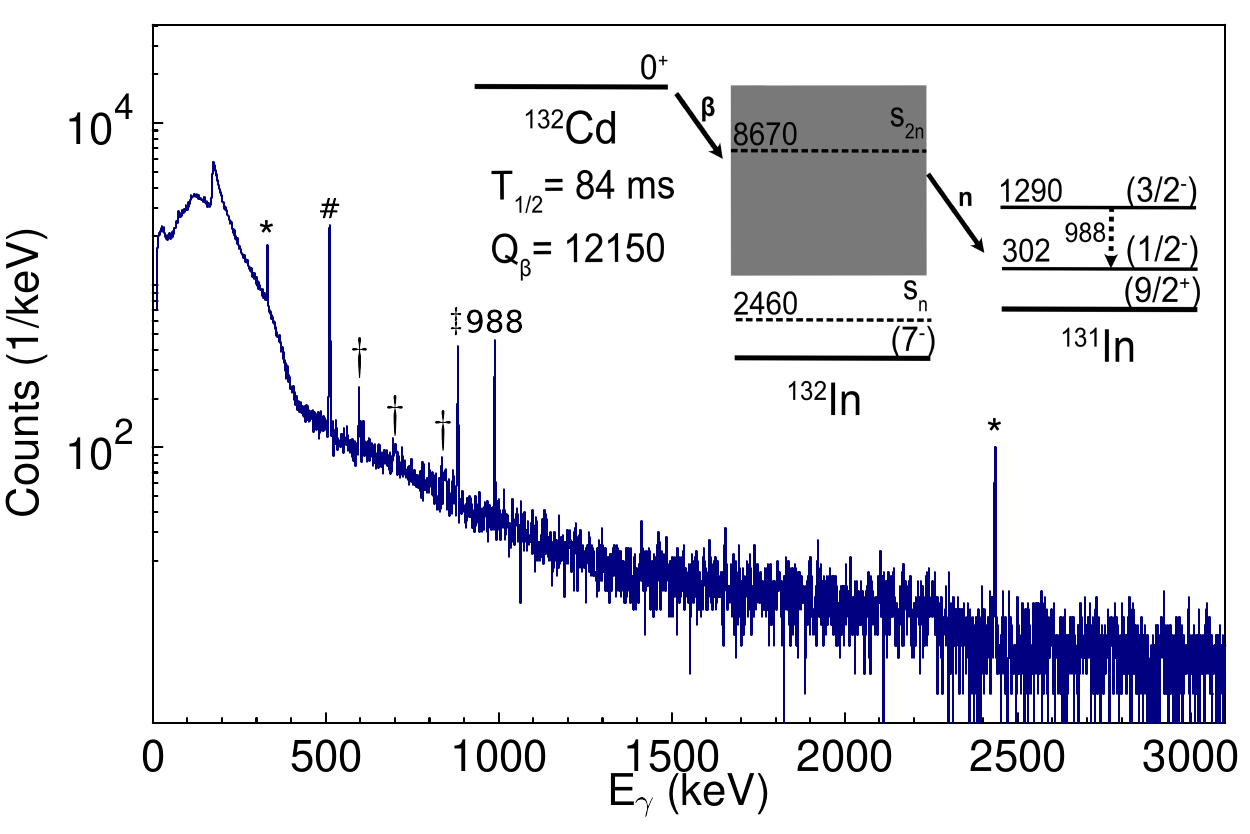}
  \caption{Gamma spectrum following the decay of $^{132}$Cd. We observe no gamma transitions in $^{132}$In and just one transition in $^{131}$In at 988 keV, populated after beta-delayed neutron emission \cite{Taprogge2014}. We observe two transitions following the decay of $^{131}$In (*), and a variety of background lines corresponding to neutron scattering in Ge ($\dagger$), $^{84}$Rb contamination from a previous experiment ($\ddagger$), and the 511 keV positron annihilation line (\#). The top right inset shows a schematic of the $^{132}$Cd beta-decay.}
  \label{fig1}
  \end{center}
\end{figure}

{\it Experiment.} The nucleus of interest, $^{132}$Cd, was produced at the ISOLDE facility at CERN using the Isotope Separation On-Line (ISOL) technique. A 1.4 GeV proton beam of 2$\mu$A intensity provided by the Proton Synchrotron Booster (PSB) induced fission in a UC$_x$ target heated to 2200 degrees Celsius located at the ion source assembly at the HRS (High Resolution Separator) beam-line. We selectively ionized $^{132}$Cd using the Resonance Ionization Laser Ion Source (RILIS) \cite{Mishin1993}. The enhanced ionization rate afforded by the use of RILIS allowed for the use of the ``cold" (room temperature) ion source transfer line in order to suppress highly volatile iodine and cesium  mass 132 isobars. These two nuclides are at the high mass peak of $^{238}$U fission yields, resulting in a potential source of background radioactivity in the $^{132}$Cd beam. The combination of laser ionization, cold transfer line, and high resolution magnetic separation resulted in an almost complete suppression of background contaminants (see the gamma spectrum in Fig. \ref{fig1}).

After isobaric separation at the HRS, a transfer beam-line delivered $^{132}$Cd ions to the ISOLDE Decay Station (IDS) for decay measurements. We implanted the 30-keV ions into a moving tape located in the middle of the $\beta$-$\gamma$-neutron counting setup. The PSB operates in pulsed mode, delivering proton bunches 50 ms long, with 1.2 or 2.4 s between packets. We operated the tape in the take-away mode. It consisted of a 200 millisecond grow-in phase, followed by 400 millisecond decay cycle, and 500 ms transport time to the shielded section of the tape. The complete cycle finished before the next proton bunch is delivered. We observed a total number of $2.2\times10^5$ decays.
The ISOLDE Decay Station consists of four high-purity Ge (HPGe) clover detectors used to detect gamma and x-ray radiation. The photo-peak efficiency of the array was 3.2\% at 1 MeV. We measured the energies of the neutron using the Versatile Array of Neutron Detectors at Low Energy (VANDLE) \cite{Peters2016}. The configuration used here consisted of 26 3$\times$6$\times$120 cm$^3$ plastic scintillator modules capped by photomultiplier tubes at both ends. The VANDLE modules were arranged in an arc of 104.5 cm radius facing the implantation spot. Two plastic scintillators surrounding the implantation point detected beta particles, providing a start signal for VANDLE with 80\% efficiency.
The acceptance of the VANDLE array, taking into account shadowing from the support frame, was  12.0\% of 4$\pi$, with a total detection efficiency 7.8\% at 1 MeV.

\begin{figure}
  \begin{center}
  \includegraphics[width=3.5in]{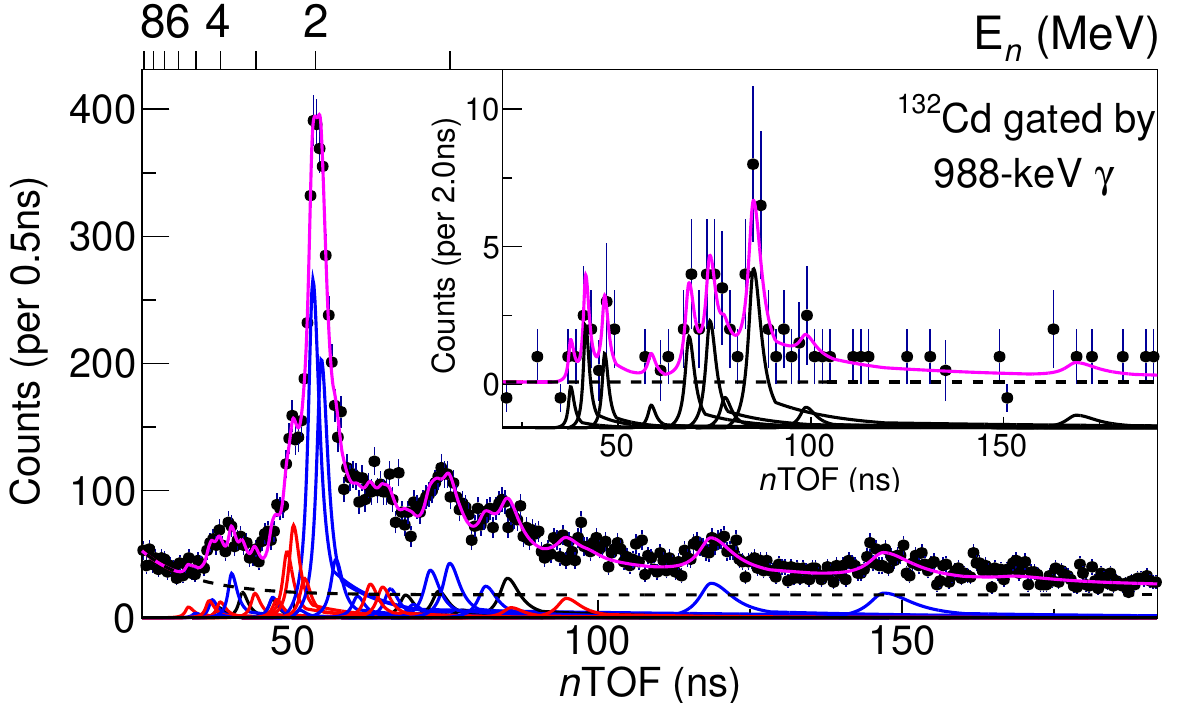}
  \caption{Beta-delayed neutron time-of-flight spectra from the decay of $^{132}$Cd. The inset shows the neutron time-of-flight spectrum in coincidence with the 988 keV gamma line in $^{131}$In (see the inset in Fig \ref{fig1}). Similarly to our observation in the decay of $^{133}$In \cite{Xu2023}, a large fraction of the neutron intensity concentrates in a peak at around 2 MeV. Both spectra were fitted simultaneously using a self-consistent procedure that includes all possible decay channels in $^{131}$In (see text for more details). The black transitions correspond to neutron emission populating the 1290 keV 3/2$^-$ state, the blue to the first excited state (1/2$^-$), and the red to the 9/2$^+$ ground state. The dashed line represents the background arising from the gamma-flash and uncorrelated events. Finally, the pink line corresponds to the sum of all transitions and background.   }
  \label{fig2}
  \end{center}
\end{figure}


Figure \ref{fig1} shows the beta-gated gamma spectrum following the decay of the implanted beam. The most prominent lines correspond to: the 988 keV  gamma transition in $^{131}$In observed by Taprogge et al. \cite{Taprogge2014}; $^{131}$Sn (331 and 2433 keV, *);  one gamma line from long lived $^{84}$Rb implanted in the chamber in a previous experiment (881 keV, $\ddagger$); and the positron annihilation line at 511 keV (\#). Daggers ($\dagger$) mark several lines identified as inelastic neutron scattering on  $^{74}$Ge in the high purity germanium gamma detectors \cite{Baginova2018}. We did not observe any of the transitions in $^{132}$In previously identified in the decay of $^{133}$Cd \cite{Jungclaus2016}, nor any transition from the decay of $^{132}$In in $^{132}$Sn \cite{Fogelberg1994}. Our gamma spectrum suggests the neutron branching ratio  of $^{132}$Cd is close to 100\%. A recent measurement at the RIBF (Rare Ion Beam Facility) facility, RIKEN, using $^{3}$He proportional counters reports a P$_{n}$ of 100\% \cite{Phong2022}, supporting our observation.

Figure \ref{fig2} shows the singles and 988 keV gamma ($^{131}$In) gated neutron spectra. The most prominent feature is a large peak at 54 ns. This indicates a large fraction of the neutron emission is concentrated at energies around 2.0 MeV. We obtained the neutron energy distribution using the fitting procedure we developed for $^{133}$In \cite{Xu2023}. First, the response of the VANDLE detector to mono-energetic neutrons was obtained from a GEANT4 Monte Carlo simulation of the array, including all detectors and support structures. Then, the response function was calibrated using the beta-delayed neutron emission of $^{17}$N \cite{Ohm1976} and $^{49}$K \cite{Carraz1982}.

\begin{figure}
  \begin{center}
  \includegraphics[width=3.5in]{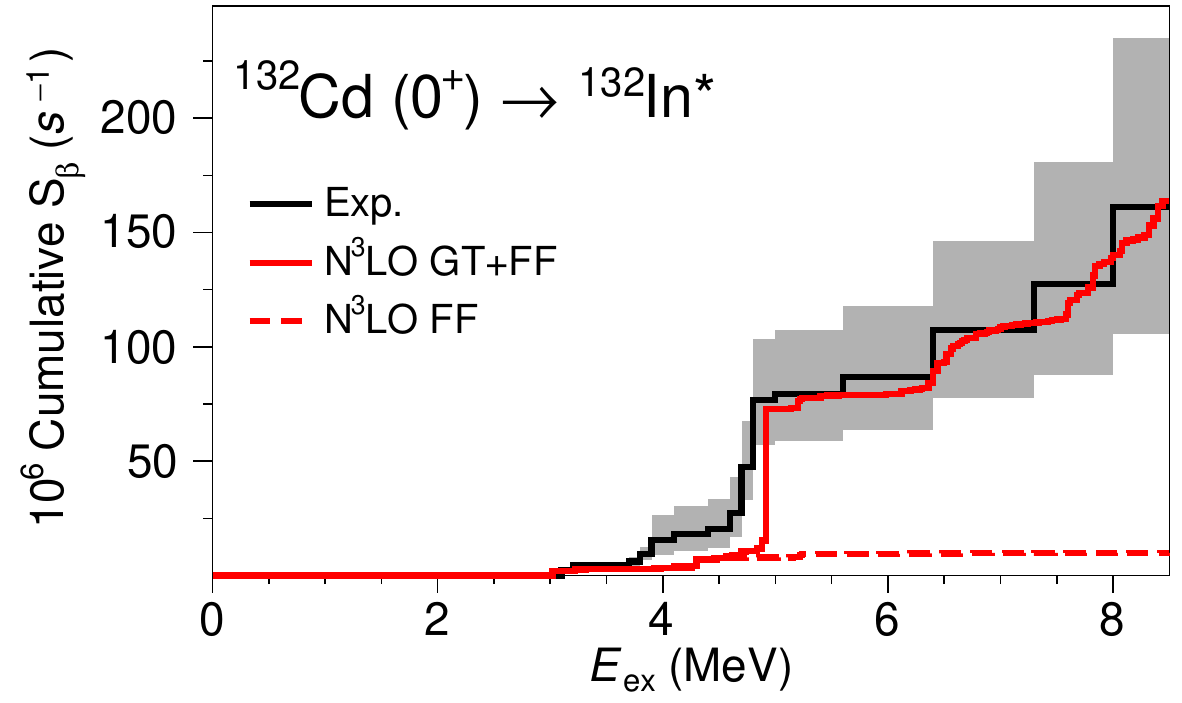}
  \caption{Cumulative decay strength as a function of the excitation energy in $^{132}$In. We obtained the strength using the intensities from the fit of the neutron time-of-flight spectrum (Fig. \ref{fig2}), the total number of $^{132}$Cd decays (2.2 10$^5$), and assuming a neutron branching ratio of 100\%. The large jump in strength at around 5 MeV corresponds to the most intense neutron transitions observed at 2 MeV in Fig. \ref{fig2}. The red line corresponds to the total decay strength calculated using the LSSM with N$^3$LO \cite{Xu2023}, while the dashed red line is just the FF strength. The model reproduces very well the experimental decay strength, further supporting its consistency across $N=82,84$ nuclei with $Z<50$.}
  \label{fig3}
  \end{center}
\end{figure}

{\it Discussion.} We obtained the cumulative decay strength distribution from the intensities obtained in the fit, the total number of $^{132}$Cd decays, and assuming a total P$_{n}$=100\%, as estimated here and by Phong et al. \cite{Phong2022}(see black line in Fig. \ref{fig3}). From the intensity of the two largest neutron transitions at 2 MeV (see Fig. \ref{fig2} in blue) populating the first 1/2$^-$ state in $^{131}$In (E$_x$=0.3 MeV) and the neutron separation energy (S$_n$=2.46 MeV \cite{Wang2021}), we deduce that the largest GT transitions correspond to two states in $^{132}$In at 4.8 and 4.9 MeV. They have B(GT) values of 0.14(2) and 0.18(2) respectively (log(ft)=4.65(7) and 4.54(6) ), well within the range of allowed GT transitions populating 1$^{+}$ states. 31(6)\% of the strength below the main transition corresponds to log(ft) values larger than 5, which are therefore possible candidates to be First Forbidden (FF) transitions.  All other transitions above 5 MeV have log(ft) values smaller than five and we propose they correspond to GT transitions to 1$^+$ states in $^{132}$In.

The red line in Fig. \ref{fig3} corresponds to the LSSM calculation of the cumulative beta-decay strength S$_{\beta}$ (for the GT transitions, S$_{\beta}$=B(GT)/6144 and quenching factor q=0.6) of $^{132}$Cd decay into $^{132}$In using N$^3$LO \cite{Entem2003}, and the dashed red line to the FF strength. The extended model space encompassed the $\pi$1p$_{1/2}$0g$_{9/2}$0g$_{7/2}$1d$_{5/2}$1d$_{3/2}$2s$_{1/2}$-$\nu$0g$_{7/2}$1d$_{5/2}$1d$_{3/2}$2s$_{1/2}$0h$_{11/2}$1f$_{7/2}$ single particle orbitals (see Z. Xu et al. \cite{Xu2023} for more details). Our LSSM calculation predicts that 17.1\% of the strength populates states through FF transitions, within two sigma of the experimental value. In our calculation, the main GT transition corresponds to a single 1$^+$ state at 4.9 MeV, with a transition strength of 0.35, matching well the observed B(GT)=0.32(3) of the two largest transitions combined. The neutron configuration of the 1$^+$ state is dominated ($>60\%$) by the neutron 0g$_{7/2}^{-1}$1f$_{7/2}^{+2}$ configuration, as expected from the large transition strength of the $\nu$g$_{7/2}\rightarrow\pi$g$_{9/2}$ transformation.

\begin{figure}
  \begin{center}
  \includegraphics[width=3.5in]{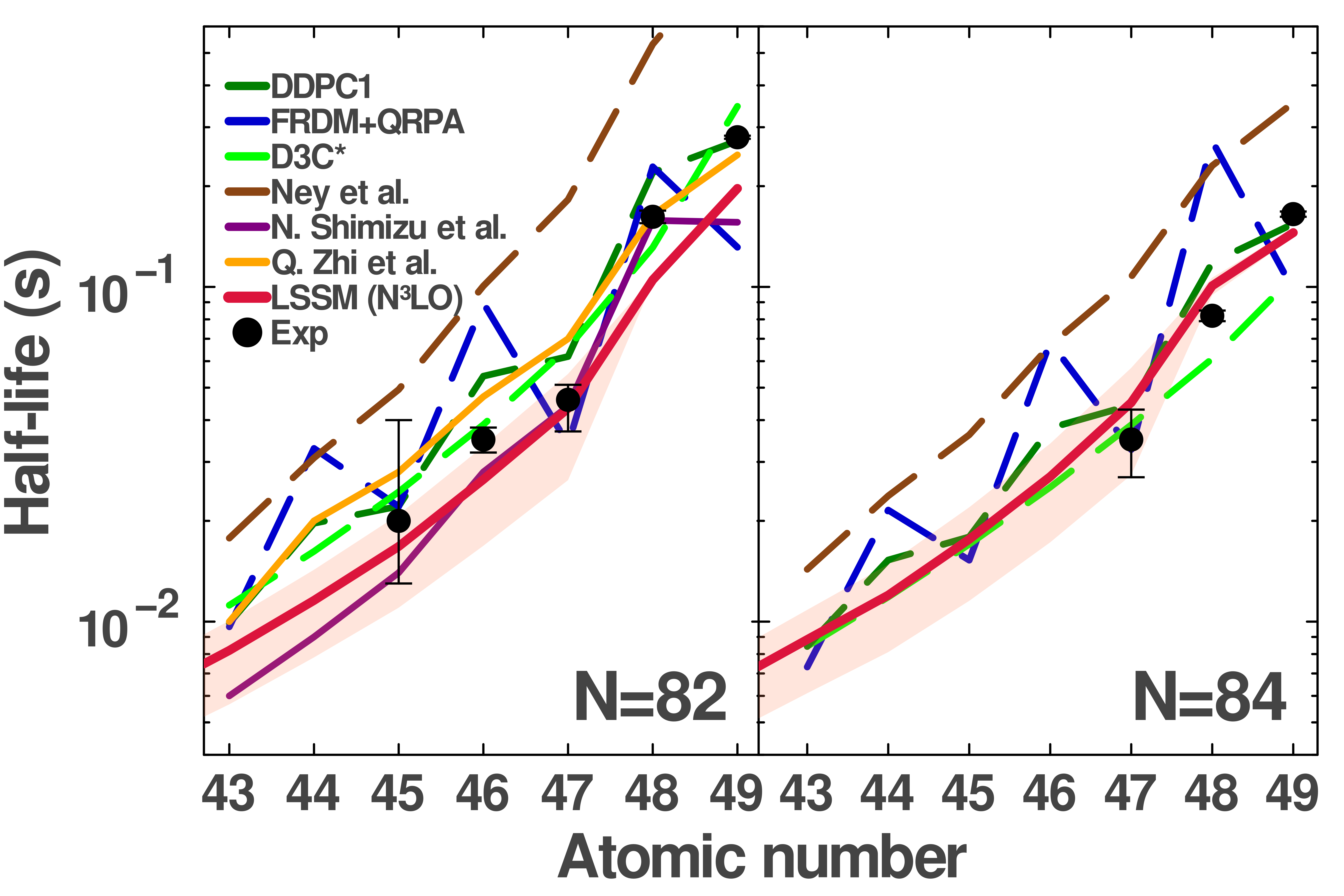}
  \caption{Left: Decay half-lives of $N=82$ isotones (\cite{nudat2000}) compared with our calculation using LSSM with N$^3$LO (red). The uncertainty band in light red corresponds to changing the unknown Q$_{\beta}$ values by $\pm$0.5 MeV. We include calculations using FRDM-QRPA \cite{Moller2019} (dashed dark blue),  D3C*  \cite{Marketin2016} (dashed green), DDPC1  \cite{Ravlic2025} (dashed dark green), Ney et al. \cite{Ney2020} (dashed brown), and shell model calculations from Shimizu et al. \cite{Shimizu2021} (purple) and Zhi et al. \cite{Zhi2013} (orange). Right panel: same for $N=84$ isotones. Note that prior to our model no other LSSM calculations were available. In both cases we see that experimental values are well reproduced by models except of Ney et al. and FRDM-QRPA, the latter being the leading model used in {\it r}-process calculations \cite{Mumpower2016,Mumpower2024}. The ``see-saw" behavior of FRDM-QRPA half-lives is driven by their treatment of pairing in their mass and Q$_{\beta}$ calculations, resulting in larger half-lives of the important even-even nuclei \cite{Mumpower2016}}
  \label{fig4}
  \end{center}
\end{figure}

\begin{figure}
  \begin{center}
  \includegraphics[width=3.5in]{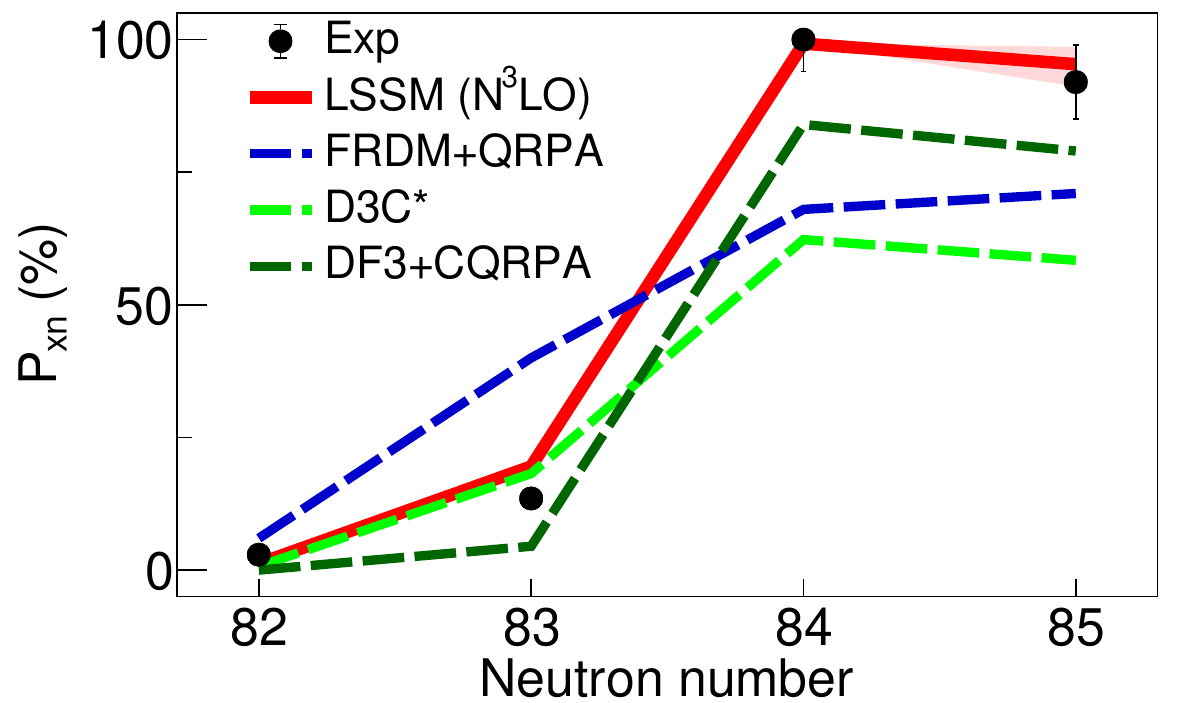}
  \caption{Beta-delayed neutron branching ratios of $N=82$ to $N=85$ cadmium isotopes \cite{nudat2000}, compared to the model using LSSM with N$^3$LO (red). For comparison, we show neutron branching ratios from FRDM-QRPA \cite{Moller2019} (dashed teal), DF3-CQRPA \cite{Borzov2016} (dashed dark green), and D3C* \cite{Marketin2016} (dashed green). While the LSSM with N$^3$LO calculation reproduces well all neutron branching ratios, all ``global models" substantially under-predict $^{132,133}$Cd.}
  \label{fig5}
  \end{center}
\end{figure}

To further test the LSSM with N$^3$LO, we calculated the decay half-lives, including both GT and FF transitions, of $N = 82$ and $N = 84$ isotones between technetium ($Z = 43$) and indium ($Z = 49$), see Fig. \ref{fig4} . For $^{131,133}$In and $^{130,132}$Cd, the Q$_{\beta}$ is taken from experiment \cite{Wang2021,Izzo2021}. For the rest ($Z \le 47$) that do not have measured atomic masses, we used the calculated Q$_{\beta}$ from LSSM. An error band corresponding to varying Q$_{\beta}$ by ±0.5 MeV was added to those nuclei to account for the uncertainty in their masses. We chose these two isotonic chains due to their impact in the {\it r}-process abundance pattern, as calculated by Mumpower et al. \cite{Mumpower2016}. We predict that the low-energy GT strength built on the microscopic $\nu g_{7/2}\rightarrow\pi g_{9/2}$ transition dominates the beta decay of all isotones below $Z = 50$. We included for comparison calculations using ``global models" FRDM-QRPA \cite{Moller2019}, D3C* \cite{Marketin2016}, DDPC1 \cite{Ravlic2025}, and the model by Ney, Engel, Li, and Schunck \cite{Ney2020}. For the $N = 82$ isotones, we also include previous shell-model calculations from Shimizu et al. \cite{Shimizu2021} and Zhi et al. \cite{Zhi2013}. Experimental half-lives are well reproduced by all models with the exception of the model by Ney et al., and even-even nuclei in FRDM-QRPA (driven by their model of nuclear pairing when calculating their masses and Q$_{\beta}$ values). All even-even nuclei shown for both $N=82$ and $N=84$ isotones are waiting points of the {\it r}-process, and all current {\it r}-process models use FRDM-QRPA predictions for unknown half-lives \cite{Mumpower2016,Mumpower2024}, decreasing their accuracy. Our model systematically predicts shorter half-lives than the FRDM-QRPA ``global" model for even-$Z$ parents, with similar results to D3C* and DDPC1. However, the FF transitions represent a much smaller fraction of the overall decay strength along N = 84, ranging between 5\% in $^{128}$Ru (31\% in D3C*, 60\% in DDPC1) and 26\% in $^{133}$In (48\% in D3C*, 75\% in DDPC1). The experimental value of $^{133}$In is determined to be 33\% \cite{Xu2023}, more consistent with our calculation. The calculations agree with all known half-lives reasonably well. Due to the difficulty in modeling the interaction between nucleons in multiple major shells, calculating beta decays with LSSM had not been available beyond $N=82$. 
We note the half-lives of $N=84$ even-even $^{128}$Ru and $^{130}$Pd have not yet been measured. In these two cases, we predict mostly pure GT decays ($\approx$95\%) with half-lives about two times shorter than the generally accepted FRDM+QRPA values \cite{Moller2019}. Although D3C* gave similar half-lives as the LSSM in these particular cases, their predictive power in general is undermined by unrealistically strong FF strengths as shown above. 

We also used our LSSM with N$^3$LO to calculate the neutron branching ratios of cadmium isotopes between $N=82$ and = 85. Figure \ref{fig5} shows the LSSM calculation  compared with ``global models" FRDM+QRPA \cite{Moller2019}, DF3-CQRPA \cite{Borzov2006}, D3C* \cite{Marketin2016}, and recent experimental values by Phong et al. \cite{Phong2022}. We note Phong et al. revised the neutron branching ratio of $N=84$ $^{132}$Cd from 60\% \cite{Hannawald2005} to 100\%, consistent with Taprogge et al. \cite{Taprogge2014} and the present result. Our LSSM calculations reproduce well all experimental values, including the neutron branching ratios of $N=84,85$ $^{132,133}$Cd, which are consistently under-predicted by all other models. We propose that this discrepancy arises from the prediction in ``global models" that a larger fraction of FF strength populates bound states compared to the LSSM calculations. For example, the large D3C* FF strength mentioned above \cite{Marketin2016} results in under-predicting the neutron branching ratio of $^{132}$Cd by a factor of two, while still being able to reproduce its half-life. This offers a clear warning against using a single experimental observation to validate theoretical calculations. On the other hand, the good agreement between the LSSM calculation of the neutron branching ratio and the experimental decay strength in $^{132}$Cd, presented here, and for $^{133}$In \cite{Xu2023} suggests that FF transitions play a smaller role than it had been proposed since the seminal work by M\"oller et al. \cite{Moller2003}.

\begin{figure}
  \begin{center}
  \includegraphics[width=3.5in]{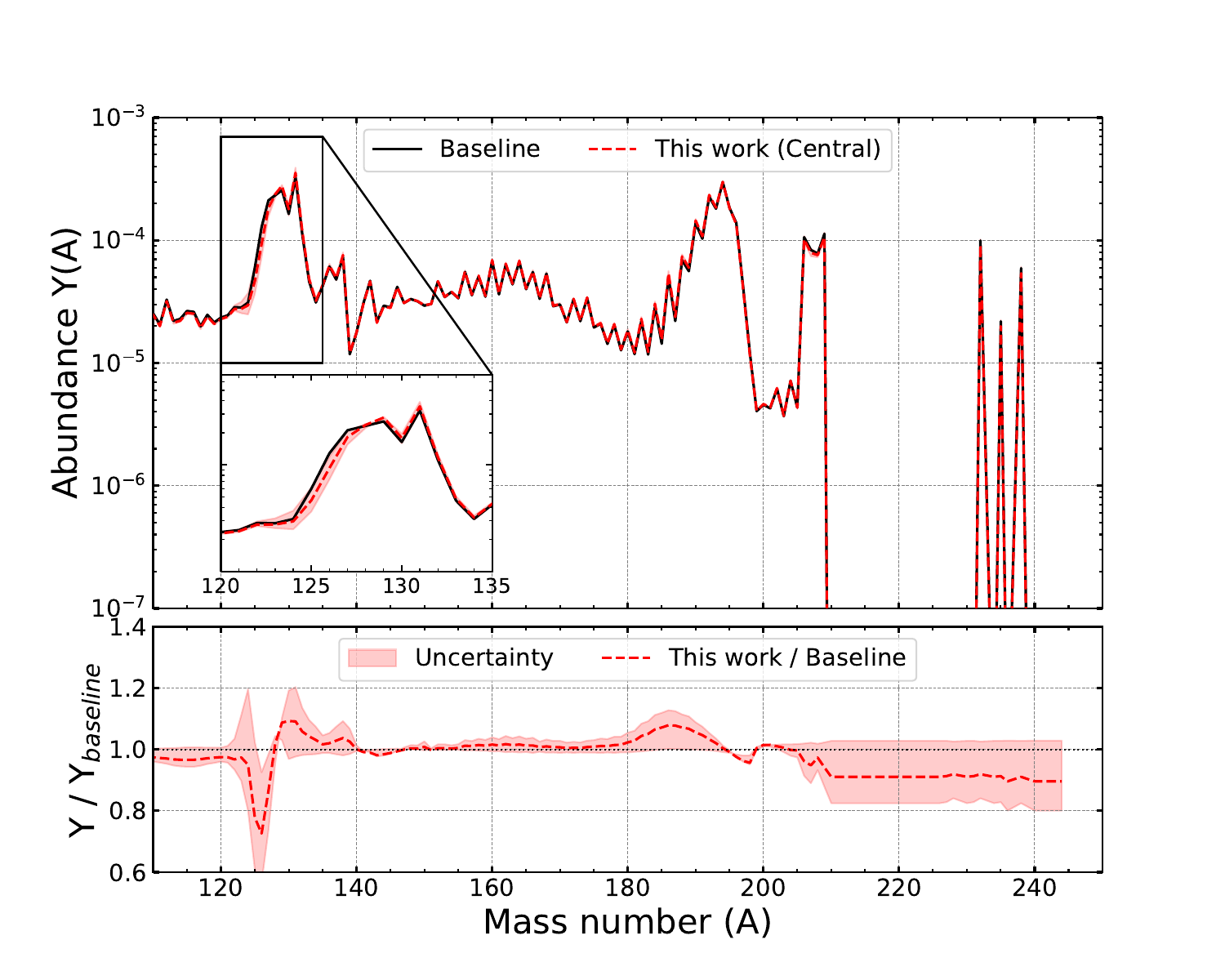}
  \caption{Top: {\it r}-process abundance calculation for a neutron star merger with robust fission recycling \cite{Mumpower2024} (black). In red we show the same calculation with  the half-lives and neutron branching ratios of $N=82$, 84 isotones between $Z=42$ and $Z=49$ replaced with our LSSM results. Please note the differences appear small due to the logarithmic scale. Bottom: ratio of the abundances using our LSSM to the baseline. The new shorter half-lives  deplete the second abundance peak at $A<130$, generating a ``hold up" on the $A>130$ side (see the inset in the top panel). More importantly, this effect is propagated to heavier nuclei in the third abundance peak (at $A\sim190$) and actinides (see text for more details). } 
  \label{fig6}
  \end{center}
\end{figure}

Finally, we investigated the effect of our newly calculated half-lives and neutron branching ratios on {\it r}-process abundance calculations. Figure \ref{fig6} shows the abundance pattern using a trajectory that represents the dynamical ejecta of a neutron star merger (with robust fission recycling) \cite{Mumpower2024}. The red line corresponds to the same calculation where the half-lives and neutron branching ratios for $N=82,84$ isotones between $Z=42$ and $Z=49$ are replaced with the values from our LSSM calculations. The red band corresponds to the uncertainty due to unknown Q$_{\beta}$ values, which was estimated by varying the central value by $\pm0.5$ MeV, as mentioned above.  We see that the new half-lives and neutron branching ratios may change the second abundance peak (at $A\sim130$) by 20\%, although still within present uncertainties. More interestingly, the abundances of the third peak (at $A\sim190$) and actinides are also modified by $\sim$10\%. The new half-lives and neutron branching ratios speed up the flow of material through the left side of the second peak, i.e. the final abundances are decreased relative to baseline. Next, the material is held up closer to the peak, at $A=129$, 130 and 131, as shown in the inset of Fig. \ref{fig6}. The ``hold up" propagates downstream to heavier nuclei in the rare earths and all the way to the actinides. The ``hold up" of material therefore decreases the amount of material that makes it to the actinides by $\sim$10\%. We propose this effect is driven by our LSSM being grounded in a more robust $N=82$ shell closure than ``global" models. 

 {\it Conclusions.} We measured the $^{132}$Cd beta-delayed gamma and, for the first time, neutron emission.   We used the time-of-flight technique to measure the neutron intensity distribution as a function of energy and obtained the GT strength populating unbound states in $^{132}$In. We note that, due to the estimated neutron branching ratio of 100\%, our GT strength distribution is likely complete up to the high-energy efficiency limit of our neutron detector. We performed Large Scale Shell Model calculations of the GT strength using the N$^3$LO interaction, resulting in an excellent agreement with our experimental distribution. The LSSM model distinguishes itself, when compared to  calculations using ``global" models, by a smaller fraction of the strength going to FF transitions feeding bound states. This, in turn, results in  predicting larger neutron branching ratios that better match known experimental values. The model also reproduces well the known half-lives of $N=82,84$ isotones between technetium and indium, as well as the $^{133,134}$In GT strength distributions previously presented in \cite{Xu2023,Heideman2023}. We utilized the half-lives and neutron branching ratios calculated using our LSSM in a {\it r}-process abundance calculation of a neutron star merger. We observe that our new half-lives and neutron branching ratios may make up to a 20\% difference with respect to the baseline abundance calculation across the second and third abundance peaks, as well as the actinides. Considering all these results together, we posit that our LSSM with N$^3$LO, anchored by our measurement of the $^{132}$Cd decay strength, is robust and applicable to the $Z<50$ and $N\geq82$ region of the chart of nuclei, of great interest to understand the nuclear structure of nuclei relevant to the astrophysical {\it r}-process.

We acknowledge the support of the ISOLDE Collaboration and technical teams. This project was supported by the European Unions Horizon 2020 research and innovation programme Grant Agreements No. 654002 (ENSAR2), by the Office of Nuclear Physics, U.S. Department of Energy under Award No. DE-FG02-96ER40983 (UTK) and DE-AC05-00OR22725 (ORNL), by the National Nuclear Security Administration under the Stewardship
Science Academic Alliances program through DOE Award No. DE-NA0002132. RL acknowledges the Romanian IFA grant CERN/ISOLDE and Nucleu project No. PN 23 21 01 02. This work was supported by the United Kingdom Science and Technology Facilities Council through grant numbers ST/P004598/1, ST/V001027/1 and ST/Y000242/1. MPS acknowledges the European Union's HORIZON Programme under the Grant Agreement No. 101212216 (RADESO). LMF acknowledges support from the Spanish MCIN/AEI/10.13039/501100011033 via project number PID2021-126998OB-I00. AIM acknowledges the support by the Spanish Ministerio de Ciencia, Innovacion y Universidades (CNS2023-144871 and PID2023-150056NB-C41), Generalitat Valenciana (CISEJI/2022/25) MJGB and OT acknowledge project PID2022-140162NB-I00 financed by the Spanish Funding Agency. The LSSM calculations
were carried out by KSHELL \cite{Shimizu2019}.



%


\end{document}